\documentclass[12pt, a4paper, twoside]{article}
\usepackage[utf8]{inputenc}
\usepackage[english]{babel}
\usepackage[a4paper, left=2cm, right=2cm, top=1.8cm, bottom=1.8cm]{geometry}
\usepackage{amsmath}
\usepackage{amssymb}
\usepackage{graphicx}
\usepackage{float}
\usepackage{wrapfig}
\usepackage{pdfpages}
\usepackage{caption}
\usepackage{cite}	
\usepackage{appendix}
\usepackage{pdfpages}
\usepackage{url}
\usepackage{slashed}
\usepackage{authblk}

\title{Detecting solar chameleons through radiation pressure}

\date{}

\begin{document}

\author[1,2]{S.~Baum \thanks{sebastian.baum@cern.ch}}
\affil[1]{\footnotesize Uppsala Universitet, Box 516, SE 75120, Uppsala, Sweden}
\affil[2]{\footnotesize European Organization for Nuclear Research (CERN), Gèneve, Switzerland}
\author[3,4]{G.~Cantatore}
\affil[3]{\footnotesize Università di Trieste, Via Valerio 2, 34127 Trieste, Italy} 
\affil[4]{\footnotesize INFN Trieste, Padriciano 99, 34149 Trieste, Italy}
\author[5]{D.H.H.~Hoffmann}
\affil[5]{\footnotesize Institut für Kernphysik, TU-Darmstadt, Schlossgartenstr. 9, D-64289 Darmstadt, Germany}
\author[4,6]{M.~Karuza}
\affil[6]{\footnotesize Phys. Dept. and CMNST, University of Rijeka, R. Matejcic 2, Rijeka, Croatia}
\author[7,8]{Y.K.~Semertzidis}
\affil[7]{\footnotesize Center for Axion and Precision Physics Research (IBS), Daejeon 305-701, Republic of Korea}
\affil[8]{\footnotesize Department of Physics, KAIST, Daejeon 305-701, Republic of Korea}
\author[9]{A.~Upadhye} 
\affil[9]{\footnotesize Physics Department, University of Wisconsin-Madison, 1150 University Avenue, Madison, WI 53706, USA}
\author[2,10]{K.~Zioutas \thanks{konstantin.zioutas@cern.ch}}
\affil[10]{\footnotesize University of Patras, GR 26504 Patras, Greece}

\maketitle

\abstract{Light scalar fields can drive the accelerated expansion of the universe. Hence, they are obvious dark energy candidates. To make such models compatible with tests of General Relativity in the solar system and ``fifth force" searches on Earth, one needs to screen them. One possibility is the so-called ``chameleon" mechanism, which renders an effective mass depending on the local matter density. If chameleon particles exist, they can be produced in the sun and detected on Earth exploiting the equivalent of a radiation pressure. Since their effective mass scales with the local matter density, chameleons can be reflected by a dense medium if their effective mass becomes greater than their total energy. Thus, under appropriate conditions, a flux of solar chameleons may be sensed by detecting the total instantaneous momentum transferred to a suitable opto-mechanical force/pressure sensor. We calculate the solar chameleon spectrum and the reach in the chameleon parameter space of an experiment using the preliminary results from a force/pressure sensor, currently under development at INFN Trieste, to be mounted in the focal plane of one of the X-Ray telescopes of the CAST experiment at CERN. We show, that such an experiment signifies a pioneering effort probing uncharted chameleon parameter space.}

\paragraph{Introduction}
The standard model of cosmology, the $\Lambda$CDM model, describes the history of our universe based on Einstein's theory of general relativity (GR), cold dark matter (CDM), and a cosmological constant ($\Lambda$). Whilst being in good agreement with astronomical and astrophysical observations, it provides no explanation for the value of its parameters. Although accounting for the accelerated expansion of the universe by a cosmological constant is the simplest model, its value must be fine-tuned. Such problems of the $\Lambda$CDM model motivate modified gravity models (MOG). One such model is scalar-tensor gravity, where GR is modified by introducing a gravitationally coupled scalar field $\phi$. A general action in the Einstein frame reads: 
\begin{equation} S = \int d^4x\,\sqrt{-g}\left(\frac{\mathcal{R}}{16 \pi G} - \frac{1}{2}\left(\partial \phi\right)^2 - V\left(\phi\right) \right) + S_m\left[g^J\right], \end{equation}
where $\phi$ is the gravitationally coupled scalar field with potential $V\left(\phi\right)$, $g = \left|g_{\mu\nu}\right|$ the determinant of the metric, $\mathcal{R}$ the Ricci scalar, $G$ the gravitational constant, and $S_m$ the matter action depending on the Jordan frame metric $g^J$, which is related to $g_{\mu\nu}$ by the conformal transformation $g^J_{\mu\nu} = A^2\left(\phi\right) g_{\mu\nu}$. In general, $A\left(\phi\right)$ is allowed to be different for the various matter fields, thus giving rise to violations of the weak equivalence principle.
\par{} However, to be in agreement with GR tests in the solar system, the effects of such scalar fields must be screened on corresponding length scales. The chameleon-model \cite{Kho04-cosmo,Kho04-fields} realizes such a mechanism by rendering an effective mass of the scalar depending on the local matter density.
\par{}The equation of motion for $\phi$ derived from the action is
\begin{equation} \partial^2 \phi = V_{,\phi}\left(\phi\right) + A_{,\phi}\rho, \end{equation}
where $\rho$ is the matter density, related to the Einstein frame density $\rho^E$ and the Jordan frame density $\rho^J$ by $\rho = A^{-1}\rho^E = A^{-3}\rho^J$. The dynamics of $\phi$ are governed by the matter density dependent effective potential $V_{\text{eff}}\left(\phi\right) \equiv V\left(\phi\right) + A\left(\phi\right)\rho$.
\par{}For suitable $V\left(\phi\right)$ and $A\left(\phi\right)$, the field will have a local minimum at some value $\phi_{\text{min}} = \phi_{\text{min}}\left(\rho\right)$. The effective mass is calculated from the curvature of the effective potential at $\phi_{\text{min}}$:
\begin{equation} m^2_{\phi,\text{eff}} = V_{\text{eff},\phi\phi} = V_{,\phi\phi}\left(\phi_{\text{min}}\right) + A_{,\phi\phi}\left(\phi_{\text{min}}\right) \rho, \end{equation}
which itself is explicitly dependent on the local mass density $\rho$.
\par{}It has been shown that chameleons cannot explain the observed accelerated expansion of the universe as as true MOG effect \cite{Khoury13}, however, they can act as dark energy. In the following, we work with an inverse power law potential
\begin{equation} V\left(\phi\right) = \lambda \frac{\Lambda^{4+n}}{\phi^n}, \end{equation}
where $\Lambda$ is a mass scale. For $n \neq -4$, the $\lambda$-parameter can be absorbed into $\Lambda$. If one adds a constant term, the chameleon fields acts like a cosmological constant on large scales, hence becomes a dark energy model. Such a potential can also be understood as the first (non-trivial) order approximation of an exponential potential:
\begin{equation} V\left(\phi\right) = \Lambda^4 e^{\Lambda^{n}/\phi^n} \approx \Lambda^4\left( 1 + \frac{\Lambda^n}{\phi^n} \right). \end{equation}
Since the constant term has no effect on the dynamics of the field, we neglect it in the following and absorb $\lambda$ into $\Lambda$ for $n \neq -4$. The chameleon-matter coupling is parametrized by $A\left(\phi\right) = e^{\frac{\beta_m}{M_{\text{Pl}}} \phi}$. Our effective potential now reads:
\begin{equation} V_{\text{eff}}\left(\phi\right) = \frac{\Lambda^{4+n}}{\phi^n} + e^{\frac{\beta_m}{M_{\text{Pl}}} \phi} \rho_{m} + e^{\frac{\beta_{\gamma}}{M_{\text{Pl}}}\phi} \rho_{\gamma}, \end{equation}
where $\rho_{m}$ is the local matter density. We assume a universal chameleon-matter coupling $\beta_{m,i} \equiv \beta_m$ for simplicity. $\rho_{\gamma} = \frac{1}{4}F_{\mu\nu}F^{\mu\nu}$ is the Lagrangian density of the electromagnetic field. Whether the chameleon-photon coupling exists at tree level or arises through fermion loops is a matter of ongoing debate \cite{Brax11,Hui10}. Here, we add the coupling by hand through a term $e^{\beta_{m}/M_{\text{Pl}}\phi}\frac{1}{4}F_{\mu\nu}F^{\mu\nu}$ in the action and consider $\beta_{\gamma}$ to be independent of $\beta_m$. From this effective potential we find
\begin{equation} \phi_{\text{min}} = \left( \frac{n \Lambda^{4+n} M_{\text{Pl}}}{\beta_m \rho_m} \right)^{\frac{1}{n+1}}, \end{equation}
where we assume $\beta_m/M_{\text{Pl}}, \beta_{\gamma}/M_{\text{Pl}} \ll \phi^{-1}$ and $\rho_{\gamma} \ll \rho_m$. The effective chameleon mass reads:
\begin{equation} V_{\text{eff},\phi\phi}\left(\phi\right) = n\left(n+1\right)\frac{\Lambda^{4+n}}{\phi^{n+2}} + \left(\frac{\beta_m}{M_{\text{Pl}}}\right)^2 e^{\frac{\beta_m}{M_{\text{Pl}}} \phi} \rho_{m} + \left(\frac{\beta_\gamma}{M_{\text{Pl}}}\right)^2 e^{\frac{\beta_{\gamma}}{M_{\text{Pl}}}\phi} \rho_{\gamma} \approx n\left(n+1\right)\frac{\Lambda^{4+n}}{\phi^{n+2}}, \end{equation}
with approximations as above and neglecting terms $\mathcal{O}\left(\frac{\beta^2}{M_{\text{Pl}}^2}\right)$. For small excitations around the minimum the effective mass is:
\begin{equation} m^2_{\text{eff}} = \left(n+1\right)\frac{\beta_m\rho_m}{M_{\text{Pl}}} \frac{1}{\phi_{\text{min}}} \propto \rho_m^{\frac{n+2}{n+1}}. \label{eq:m_eff} \end{equation}
Hence, for $n>-1$ and $n<-2$ chameleons get large effective mass in dense environments which leads to the screening effect. For $-2 \leq n \leq -1$ there is no chameleon effect. For $n$ between $-1$ and $\approx 1/2$ the scaling is very rapid, leading to very strongly screened chameleons which are best probed cosmologically \cite{EotWashRes}. Thus, for this work we are primarily interested in $n \gtrsim 1/2$ and $n<-2$.
\par{} Chameleons would be produced in regions of high photon density and strong magnetic field, e.g. inside the sun. One can search for solar chameleons on Earth by detecting their radiation pressure \cite{Baker12}: In a dense medium chameleons get large effective mass. If their total energy is smaller than their effective mass in a medium they try to penetrate, they will get reflected, resulting in the equivalent of radiation pressure. In the following, we calculate the sensitivity of such an experiment in three parts:
\begin{itemize}
\item chameleon production in the sun
\item the chameleons' journey to the detector
\item detecting chameleons on Earth.
\end{itemize}

\paragraph{Chameleon production in the sun} 
Armed with our chameleon model, we can study chameleon production in the sun. We build our calculations on the previous works of Brax \emph{et al.} \cite{Baker12, Brax10, Brax12}. Photons mix with chameleons in regions of strong magnetic field, cf. the case for Axions. In the sun, strong magnetic fields are found in the so-called tachocline, the thin transition zone between the radiative core of the sun exhibiting almost solid rotation and the convective envelope rotating differentially. Here, the large shear causes strong magnetic fields \cite{Cou03}.
The conversion probability for photons of energy $\omega$ in a magnetic field $B$ traveling by a length $L$ is given by \cite{Burr09}:
\begin{equation} P_{\text{chameleon}} \left(\omega\right) = \sin^2\left(2\theta\right) \sin^2\left(\frac{\Delta}{\cos 2\theta} \right), \end{equation}
where $\Delta = \left(m_{\text{eff}}^2-\omega_{\text{pl}}^2\right) L / 4 \omega$ and $\tan\left(2\theta\right) = \frac{2 B \omega \beta_{\gamma}}{\left(m^2_{\text{eff}}-\omega_{\text{Pl}}^2\right)M_{\text{pl}}}$ is the mixing angle. $\omega_{pl}^2 = 4\pi\alpha n_e/m_e$ is the plasma frequency with $n_e$ the electron number density, $m_e$ the electron mass, and $\alpha$ the fine structure constant. If chameleons are predominantly produced in the sun's tachocline, this simplifies greatly. Assuming $\rho_{\text{tacho}} = 0.2\,$g/cm$^3$, $B_{\text{tacho}} = 30\,$T, and $T_{\text{tacho}} = 2 \times 10^6\,$K \cite{Cou03}, we find $\tan\left(2\theta\right) \sim 10^{-20}\,$eV$^2\times\frac{\beta_{\gamma}}{m_{\text{eff}}^2-\omega_{\text{Pl}}^2}$. Chameleons can only travel in the sun for an effective momentum
\begin{equation} k^2 = \omega^2 - \left( m_{\text{eff}}^2 - \omega_{\text{pl}}^2 \right) \geq 0. \end{equation}
As we will see below, stellar evolution constraints $\beta_{\gamma} \leq 10^{10}$ for large parts of the parameter space. Even for $\beta_{\gamma}$ some orders of magnitude larger we find $\tan\left(2\theta\right) = \sin\left(2\theta\right) = 2\theta$ and $\cos\left(2\theta\right) = 1$. Furthermore, we find $\Delta\left(L\right) \gtrsim 10^3 \times \left(\frac{L}{1\,\text{cm}}\right)$ where we need to compare $L$ with the photon mean free path in the tachocline $\lambda \approx 0.25\,$cm \cite{Mit92}. Since $\Delta\left(\lambda\right) \gg 2\pi$, we average $ \left\langle \sin^2 \Delta\left(\lambda\right) \right\rangle = 1/2 $. Hence, the conversion probability reads
\begin{equation} P_{\text{chameleon}}\left(\omega\right) = 2 \theta^2 = 2\left( \frac{\omega B \beta_{\gamma}}{M_{\text{Pl}} \left( m_{\text{eff}}^2 - \omega_{\text{pl}}^2 \right)} \right)^2. \label{eq:P_cham_simpl}\end{equation}
Once produced in the tachocline, the chameleons leave the sun unscathed: the interaction rate with fermions can be estimated as $\Gamma_f = n_f \beta_m^4 \frac{m_f^2}{M_{\text{Pl}}^4}$ \cite{Brax12}. Hence, interactions with protons dominate in the sun. With $n_f = \rho_{\text{sol}}/m_p$, we can estimate the mean free path of chameleons in the sun to
\begin{equation}\lambda_{\text{chameleon}} = \frac{1}{\Gamma_p} =  \beta_m^{-4} \frac{M_{\text{Pl}}^4}{m_p \rho_{\text{sol}}} \sim 10^{24} \times \left(\frac{10^{10}}{\beta_m}\right)^4 \times \left(\frac{100\,\text{g.cm}^{-3}}{\rho_{\text{sol}}}\right) \times R_{\text{sol}},\end{equation}
where $R_{\text{sol}}$ is the solar radius. Corrections can occur through quantum corrections above the chameleons' effective field theory cutoff scale $\Lambda$. The past several years have seen considerable discussions about the extent to which such corrections affect the interpretation of laboratory searches: one effect of particular concern for this work is fragmentation, by which a small number of chameleon particles can interact to produce a larger number of lower-energy chameleons. Such interactions can occur either through self-interaction terms in the potential or through interactions with photons and Fermions in the sun. Studies of fragmentation for afterglow experiments such as GammeV-CHASE have as yet not been extended to incoherent conversion in thermal systems such as the sun, although the results do demonstrate that fragmentation does not necessarily become large at energies above the cutoff \cite{Brax14}. Studies of chameleon production in the high-temperature universe during Big Bang Nucleosynthesis found large, unphysical amounts of particle production caused by ``kicks'' to the chameleon, as other particles transitioned from relativistic to nonrelativistic \cite{Erickcek14}. These results, however, also do not apply to solar production, since the sun is sufficiently stable and nonrelativistic. A study of fragmentation effects inside the sun goes beyond the scope of this work. We show below how the results of such effect can be mitigated in the experimental setup.
\par
\begin{figure}
	\includegraphics[height=7cm]{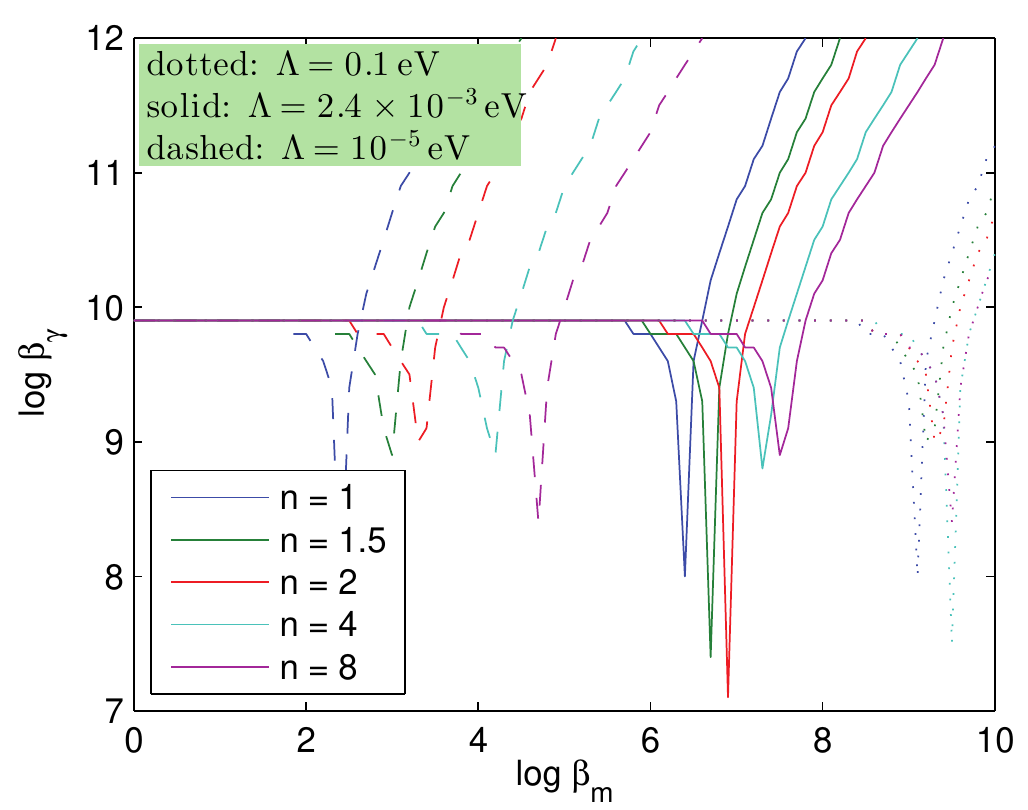}
	\includegraphics[trim = .9cm 0cm 0cm 0cm, clip, height=7cm]{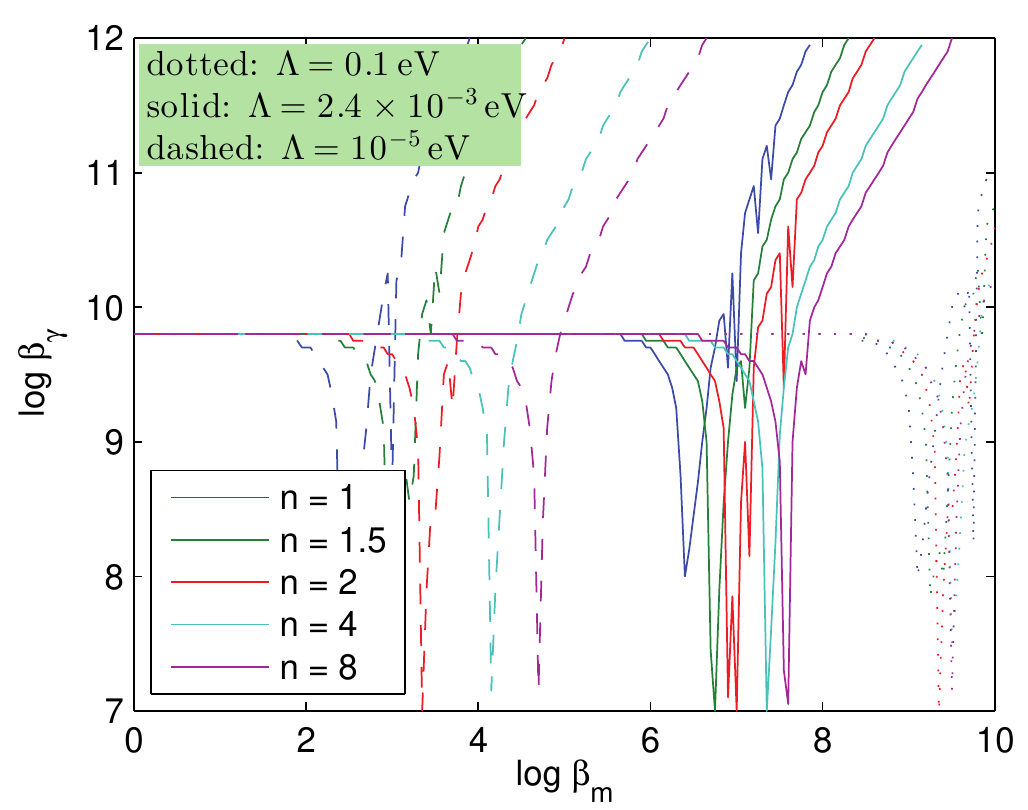}
	\includegraphics[trim = 0cm 0cm 1.85cm 0cm, clip, height=7cm]{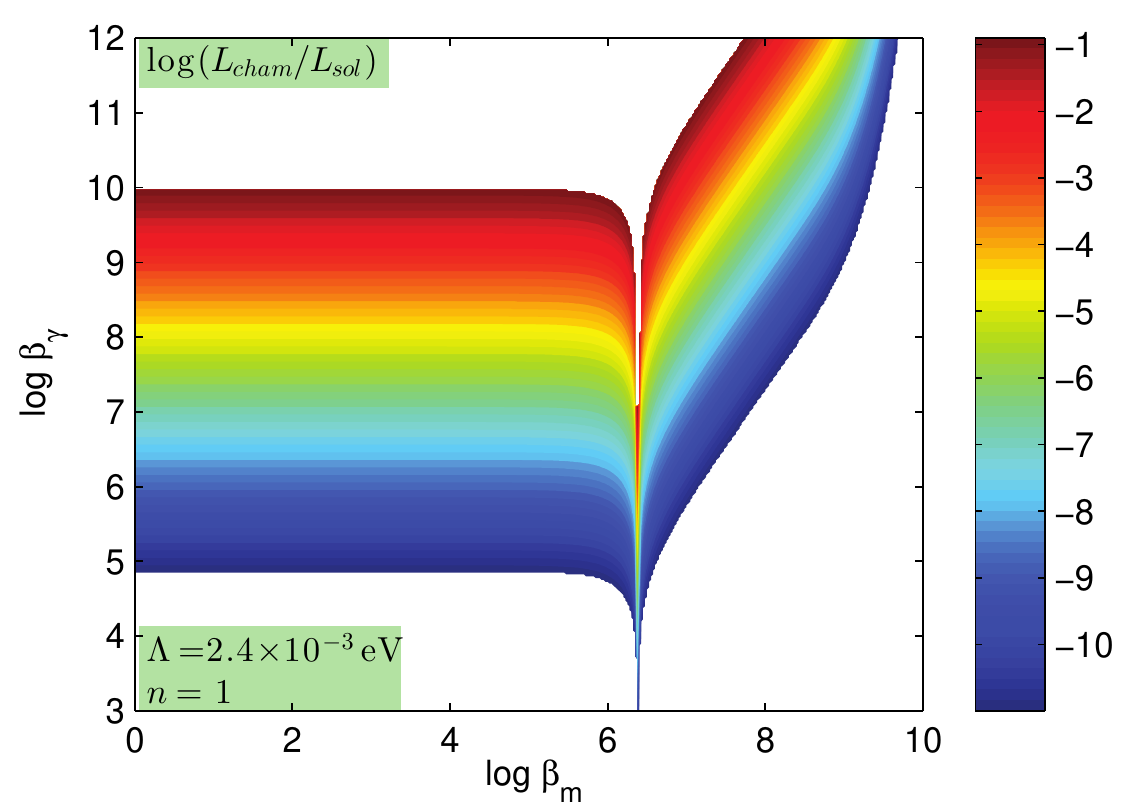}
	\includegraphics[trim= .9cm 0cm 0cm 0cm, clip, height=7cm]{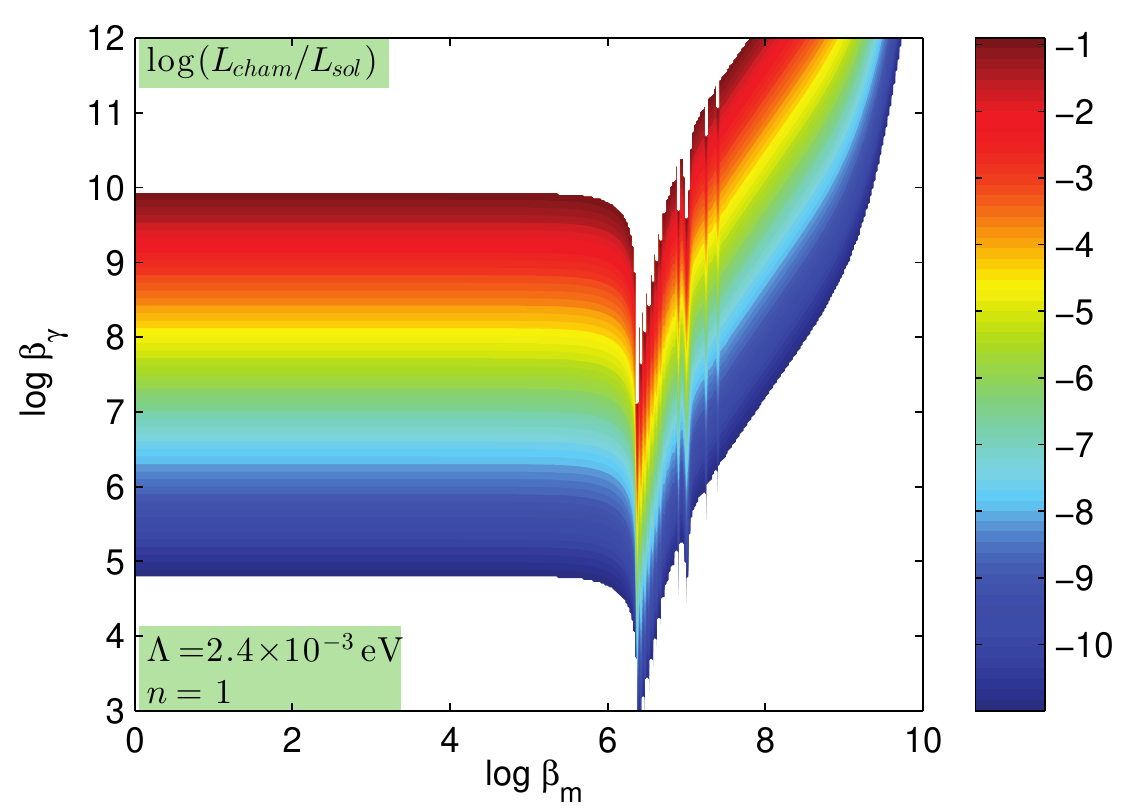}
	\caption{The top two panels show the photon coupling $\beta_{\gamma}$, for which the chameleon luminosity is 10\,\% of the visible solar luminosity, depending on the other chameleon parameters $\beta_m, n, $ and $\Lambda$. From this one can immediately calculate the chameleon luminosity for any other $\beta_{\gamma}$, since $L_{\text{cham}} \propto \beta_{\gamma}^2$, with all other parameters unchanged. The two lower panels show this scaling of $L_{\text{cham}}$ with $\beta_{\gamma}$ explicitly for the case $n = 1$, $\Lambda = 2.4\times10^{-3}\,$eV and tachocline parameters as before. Here, the white areas correspond to $L_{\text{cham}}/L_{\text{sol}} > 0.2$ (above), and $L_{\text{cham}}/L_{\text{sol}} < 10^{-11}$ (below), respectively. The leftmost panels correspond to chameleon production inside the tachocline only, with $B = 30\,$T for $R = 0.70 \ldots 0.75\,R_{\text{sol}}$. The rightmost panels correspond to a magnetic field linearily decreasing from $B = 30\,$T at $R = 0.7\,R_{\text{sol}}$ to $B = 1\,$T at the solar surface.}
	\label{fig:10percentandlumi}
\end{figure}

{} Photons perform a random walk inside the sun and their mean free path $\lambda_{\gamma} \ll R_{\text{sol}}$. Hence, traversing a shell of thickness $\Delta R$ a photon on average can convert $N = \Delta R^2 / \lambda^2$ times. Putting all of the above together, we calculate the contribution of a shell of thickness $\Delta R$ at a distance $R$ from the solar center to the chameleon spectral density at the solar surface:
\begin{equation} u_{\text{chameleon}}\left(\omega\right) = \frac{R^2}{R^2_{\text{sol}}} \times \frac{\Delta R^2}{\lambda^2} \times u_{\gamma,\text{tacho}}\left(\omega\right) \times P_{\text{chameleon}}\left(\omega\right) \times \Theta\left(k^2\right), \label{eq:cham_spec} \end{equation}
where $\Theta\left(k^2\right)$ is the Heaviside step function and the photon spectrum in the tachocline is given by 
\begin{equation} u_{\gamma,\text{tacho}}\left(\omega\right) = \frac{\omega^3}{4 \pi^4} \frac{1}{e^{\omega/T_{\text{tacho}}} - 1}. \end{equation}
Integrating eq. \ref{eq:cham_spec} over the inside of the sun, we get the solar chameleon spectrum. We consider two different cases of magnetic field in the sun: a magnetic field $B = 30\,$T inside the tachocline $R_{\text{tacho}} = 0.70 \ldots 0.75\,R_{\text{sol}}$ \cite{Cou03}, and a linearily decreasing field from $B=30\,$T at $R=0.7\,R_{\text{sol}}$ to $B=1\,$T at the solar surface \cite{Anita03}. The resulting chameleon luminosity $L_{\text{cham}}$ is shown in Figure \ref{fig:10percentandlumi}. Since the chameleon luminosity scales with $\beta_{\gamma}^2$, with all other parameters left unchanged, we can immediately calculate the luminosity for any $\beta_{\gamma}$ once we know it for one value. Since solar evolution constrains the total exotica flux from the sun to $L_{\text{exotica}}/L_{\text{sol}} \leq 0.1$, we can exclude chameleon models with stronger photon couplings than shown in the top panels of Figure \ref{fig:10percentandlumi}. This bound also justifies the approximations made for eq. \ref{eq:P_cham_simpl}.
\par{}We discuss the case $\Lambda = 2.4\times10^{-3}\,$eV, the dark energy scale, $n = 1$, and chameleon production only in the tachocline in detail: for $\beta_{m} < 10^{5.9}$ we find the chameleon luminosity $L_{\text{cham}}/L_{\text{sol}} = 0.1$ of the visible solar luminosity for $\beta_{\gamma} = 10^{9.9}$. With growing $\beta_m$ one passes a zone of enhanced production since $m_{\text{eff}} \approx \omega_{Pl}$. We find $L_{\text{cham}}/L_{\text{sol}} = 0.1$ for roughly two orders of magnitude smaller $\beta_{\gamma}$ than before, cf. the discussion of this case in \cite{Brax10, Brax12}. For even greater $\beta_{m}$ the chameleon luminosity decreases steeply, since an increasing part of the spectrum is too heavy to propagate inside the sun.
\par{}Since the chameleon production is proportional to $T^{4}B^2\rho^{-2}$, shells outside the tachocline only give small contributions to the chameleon flux in the case of the linearily decreasing magnetic field model. Only for $\beta_{m} = 10^{6.3} \ldots 10^{7.4}$ does this case differ significantly from chameleon production only inside the tachocline; then the case of enhanced photon production discussed above occurs outside the tachocline and gives rise to a low energy enhancement of the chameleon spectrum. The luminosity for different parameter sets can be read off Figure \ref{fig:10percentandlumi}.

\paragraph{The chameleons' journey to the detector}
Chameleons leaving the sun will travel to the detector unscathed, provided their energy is greater than their effective mass in whichever medium they propagate. Since the effective mass scales like $m_{\text{eff}} \propto \rho_m^\frac{n+2}{2\left(n+1\right)}$, the chameleon spectrum at the detector is cut off for $\omega_{\text{chameleon}} < m_{\text{eff}}\left(\rho_{\text{max}}\right)$ if $n > -1$ or $n < -2$, where $\rho_{\text{max}}$ is the medium of highest density the chameleons travel through. In our case, the densest medium is the window the chameleons have to penetrate when entering the vacuum chamber housing the sensor: $\rho_{\text{window}} \approx 1\,$g/cm$^{3}$. Hence, we introduce a cutoff of the chameleon spectrum $u_{\text{chameleon}}\left(\text{detector}\right) \propto u_{\text{chameleon}}\left(\text{sun}\right) \times \Theta\left(\omega-m_{\text{eff}}\left(\rho_{\text{max}}\right)\right)$.

\paragraph{Detecting chameleons on Earth}
The flux of solar chameleons reaching Earth could be detected by exploiting their direct coupling to matter $\beta_m$ with an opto-mechanical force sensor. One such sensor, called KWISP for ``Kinetic WISP detection'', is currently under development at INFN Trieste, in Italy. These sensors are typically based on a thin micro-membrane displaced from its rest position by an external force (or equivalently a pressure) applied to it. The membrane displacement is then sensed by optical means, normally using interferometry for maximum sensitivity, giving a direct measurement of the force (pressure) acting on the membrane. The sensitivity to displacements is greatly enhanced if the membrane is placed inside a high finesse Fabry-Perot (FP) optical resonator, since the finesse acts as gain factor in the sensitivity.
\par{}The working principle of such a membrane-based opto-mechanical sensor is as follows. An FP optical resonator cavity is frequency-locked to a laser beam using an electro optic feedback \cite{Cantatore1995}. The feedback acts on the laser active medium, a crystal in the most common case of a Nd:YAG laser, so that the instantaneous distance between the cavity mirrors, left ``free'' to float, is always a half-integer multiple of the laser wavelength. When the cavity is at resonance, its normal modes are not perturbed if a thin membrane, transparent to the laser wavelength, is aligned and positioned in a node of the standing intra-cavity electric field. A subsequent membrane displacement couples the membrane mechanical modes to the TEM modes of the cavity, detuning the mode proper frequencies with a typical oscillatory signature dependent on membrane position along the cavity axis \cite{Karuza12}. Once such an opto-mechanical setup has been calibrated by determining its detuning curve, it can be used to sense membrane displacements and therefore, from the membrane's mechanical characteristics, the force acting on the membrane. The real advantage in using such a complicated technique is that extremely tiny forces can be detected, as we shall see below.
\par{}The KWISP sensor now under test in the INFN Trieste laboratories employs a $5\,\text{mm}\times 5\,\text{mm}$, 
100\,nm-thick, Si$_3$N$_4$ micromembrane (made by Norcada Inc., Canada, and having $\rho = 3.2\,$g/cm$^3$) set inside an 85 mm long FP cavity in concave-concave configuration. The cavity is probed by a 1064 nm CW Nd:YAG laser. 
Optionally, the membrane may be coated with a thin metal layer \cite{Yu12}. For instance, in our calculations we consider a gold layer $20\,$nm thick. In this case, to preserve transparency, the coating will leave a clear central zone having a radius slightly larger than the beam waist (typically $\leq 1\,\text{mm}$).
\par
\begin{figure}
	\includegraphics[height=6.5 cm]{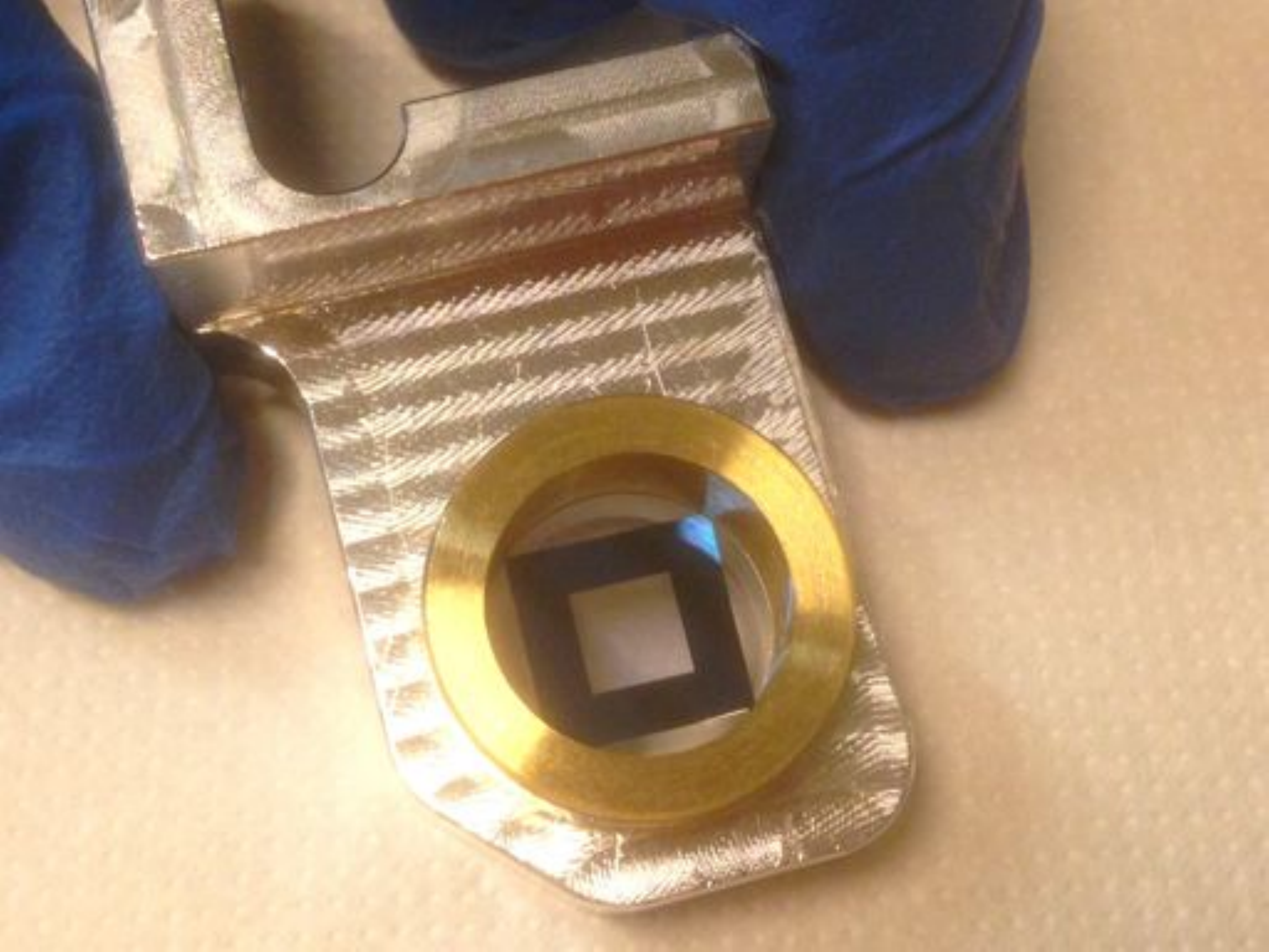}
	\includegraphics[trim = 0cm 0cm 0cm 0cm, clip, height=6.5 cm]{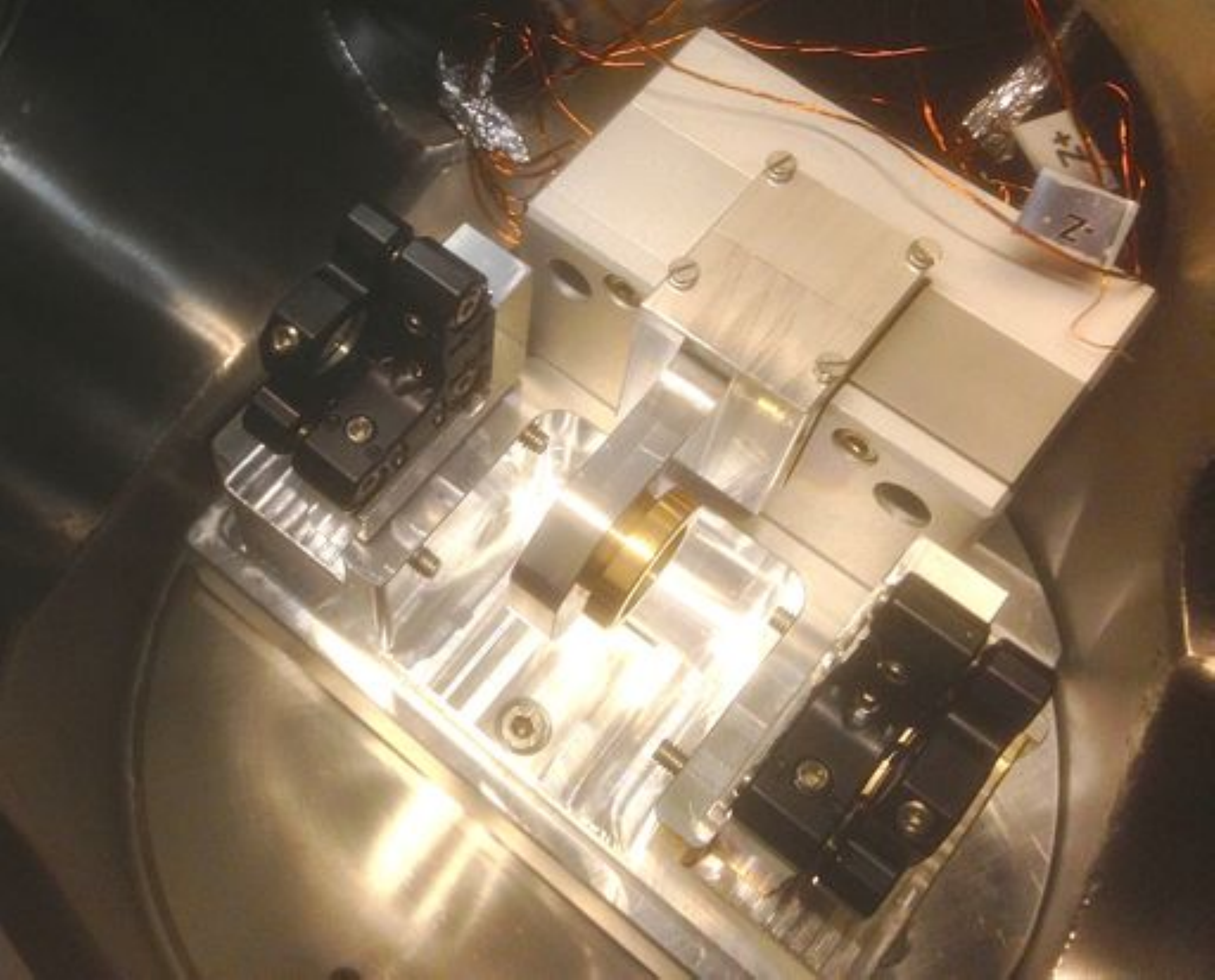}
	\caption{(At left) Photograph of the $5\,$mm$\times 5\,$mm, 100\,nm-thick Si$_3$N$_4$ micromembrane mounted inside a holder. The holder clearance is such that the membrane outer frame contacts the holder only at the corners. This avoids degrading the mechanical characteristics of the membrane itself. (At right) Photograph of the FP-membrane assembly set inside its vacuum chamber. The FP consists of two half-inch diameter, multi-layer, high-reflectivity mirrors fixed 85 mm apart (mirror mounts are visible in the picture as black structures). The membrane holder is mounted on a PZT-actuated vacuum compatible 5-axis movement stage (``Pentor" model made by Piezosystem Jena), which in turn is fixed on a machined aluminum base holding also the mirror mounts.}
	\label{fig:membrane}
\end{figure}

{}Figure \ref{fig:membrane} shows, at left, a photograph of the $5\,$mm$\times 5\,$mm membrane mounted on a holder, and, at right, a photograph of the FP-membrane assembly set inside its vacuum chamber.
The membrane can be aligned manually using a precision tilting mount, then final adjustments and movements are carried out by a PZT actuated 5-axis movement stage. The FP-membrane assembly is housed inside a vacuum chamber evacuated at $\approx 10^{-4}$ mbar or less, while laser and beam injection optics are at atmosphere. After proper alignment, the FP cavity finesse was measured to be $F \approx 60000$.

\par{}Before setting up the complete opto-mechanical sensor, in order to derive a preliminary force sensitivity figure, the membrane was inserted in a Michelson-type interferometer, corresponding to a single-pass FP. The resulting displacement sensitivity, that is the minimum membrane displacement detectable in 1 s of measuring time, was $0.18\,\text{nm}/\sqrt{\text{Hz}}$.
\par{}By inserting its mechanical characteristics into a finite element simulation program it was found that the membrane can be modeled by a simple spring with a constant of $16.6 \,\text{N}/\text{m}$ \cite{Gardikiotis}, giving a force sensitivity of $3.0\times10^{-9}\,\text{N}/\sqrt{\text{Hz}}$ (single pass).
Once the membrane is set inside the FP, this force sensitivity is directly enhanced by the finesse factor, thus the expected force sensitivity is 
$S_{force} = 5\times10^{-14}\,\text{N}/\sqrt{\text{Hz}}$. Recall that this sensitivity figure corresponds to the minimum force detectable in 1 s of measurement time, therefore measuring for, say, 10000 s would result in a factor 100 improvement in the minimum detected force level.
The KWISP force sensor is now undergoing complete characterization at INFN Trieste. Once this task is completed the sensor will be moved to the CAST experimental area at CERN for initial off-beam commissioning tests to determine the environmental compatibility of the prototype. Based on the results of the commissioning phase, the design and construction of an on-beam prototype, to be placed in the focus of the CAST X-ray telescope, will follow.
\par{}If we assume the chameleon flux $\Phi_{\text{chameleon}}$ to be $\Phi_{\text{chameleon}}=\lambda \Phi_{\text{sol}}$ and $\Phi_{\text{sol}}=1.36\,$kW/m$^2$, we need
\begin{equation} \boxed{\frac{\Phi_{\text{reflected}}}{\Phi_{\text{chameleon}}} \gtrsim 2 \times 10^{-4} \times \left(\frac{10\,\%}{\lambda} \right) \times \sqrt{\frac{100\,\text{s}}{t_{\text{meas.}}}} } \label{eq:ref_power} \end{equation}
of the total chameleon flux to be reflected by the sensor in order to detect chameleons.
\par{}The sensor membrane will reflect all chameleons with energies smaller than the chameleons effective mass in the membrane's material. Hence, the fraction of the chameleon flux reflected by the sensor is given by
\begin{equation} \frac{\Phi_{\text{reflected}}}{\Phi_{\text{chameleon}}} = \frac{\int\limits_{0}^{m_{\text{eff}}}u_{\text{chameleon}}\left(\omega\right) \times \Theta\left(\omega-m_{\text{eff}}\left(\rho_{\text{max}}\right)\right) d\omega}{\int\limits_{0}^{\infty}u_{\text{chameleon}}\left(\omega\right)d\omega}. \end{equation}
The sensitivity of the experiment can be optimized by introducing an incident angle of the chameleons on the membrane $\theta > 0$. Then all chameleons with $k_{\perp} = \omega \cos\theta \leq m_{\text{eff}}$ are reflected. On the other hand, this implies that the force on the sensor is reduced by $\cos\theta$ and the effective area of the sensor reduced by a factor $\sqrt{\cos{\theta}}$. Thus, after introducing an incident angle the fraction of the flux reflected by the sensor reads:
\begin{equation} \frac{\Phi_{\text{reflected}}}{\Phi_{\text{chameleon}}} = \left(\cos{\theta}\right)^{3/2} \frac{\int\limits_{0}^{m_{\text{eff}}/\cos{\theta}}u_{\text{chameleon}}\left(\omega\right) \times \Theta\left(\omega-m_{\text{eff}}\left(\rho_{\text{max}}\right)\right) d\omega}{\int\limits_{0}^{\infty}u_{\text{chameleon}}\left(\omega\right)d\omega}. \end{equation}
\par{}Further improvement of the experiment's sensitivity can be achieved by employing an X-Ray telescope (XRT) \cite{Baker12-2} like the ABRIXAS telescope used at CAST. Since the telescope mirrors are coated with gold and the grazing angle of the telescope is smaller than $1^{\circ}$, all chameleons reflected by the membrane will be focused by the telescope for an incident angle of the membrane $\theta<89^{\circ}$. CAST's XRT will increase the chameleon flux on the sensor by a factor 500, hence, the required flux to be reflected in eq. \ref{eq:ref_power} is reduced by 500. One can furthermore cool down the membrane below 0.3\,K, which yields another gain factor 100. Modulating the chameleon flux with a chopper would result in another sensitivity gain $>10$, depending on the structure of the low-frequency noise of the force sensor. Using recently proposed advanced optical techniques \cite{Xu13}, additional sensitivity gains could be foreseen.

\begin{figure}
	\includegraphics[width=.49\linewidth]{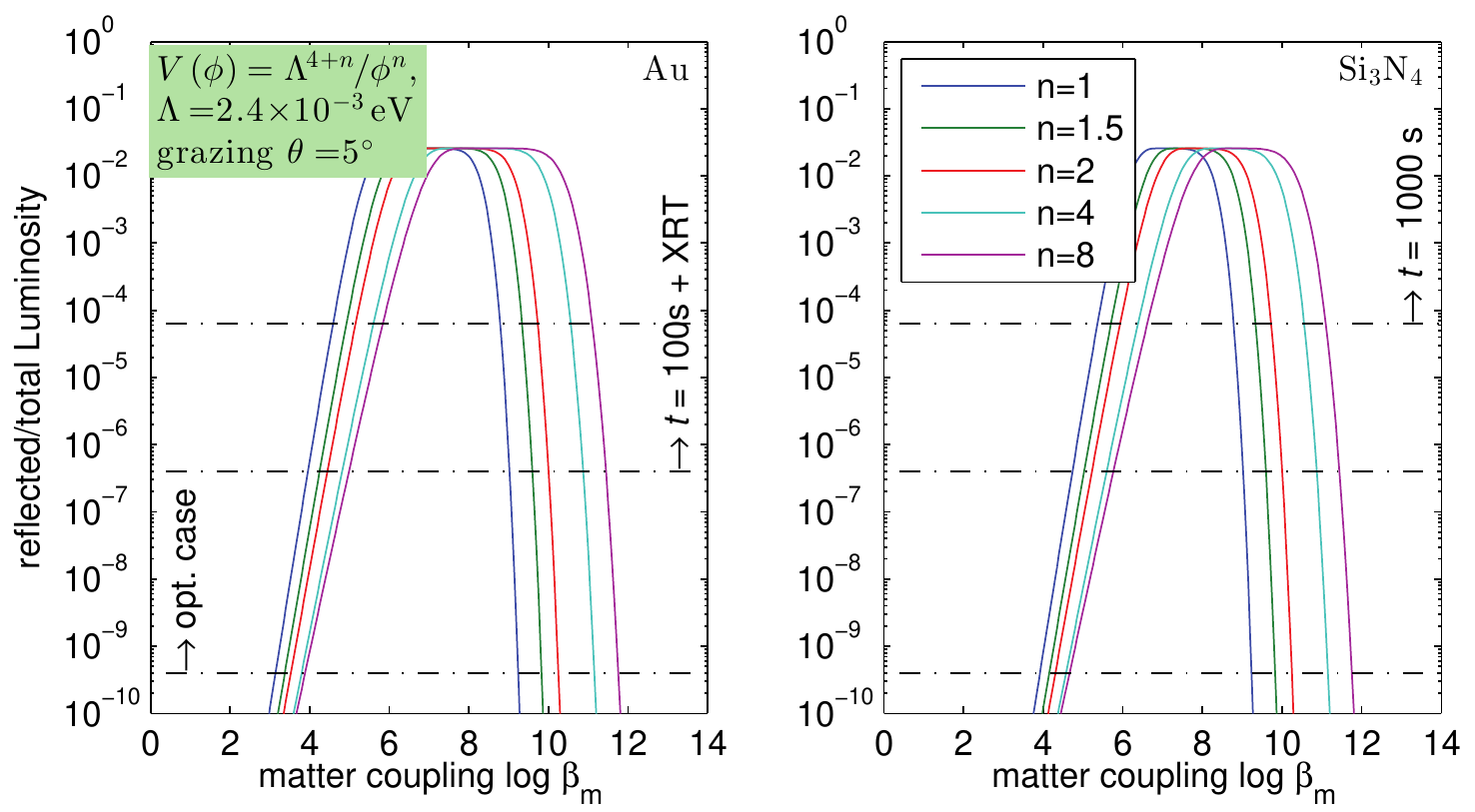}
	\includegraphics[width=.49\linewidth]{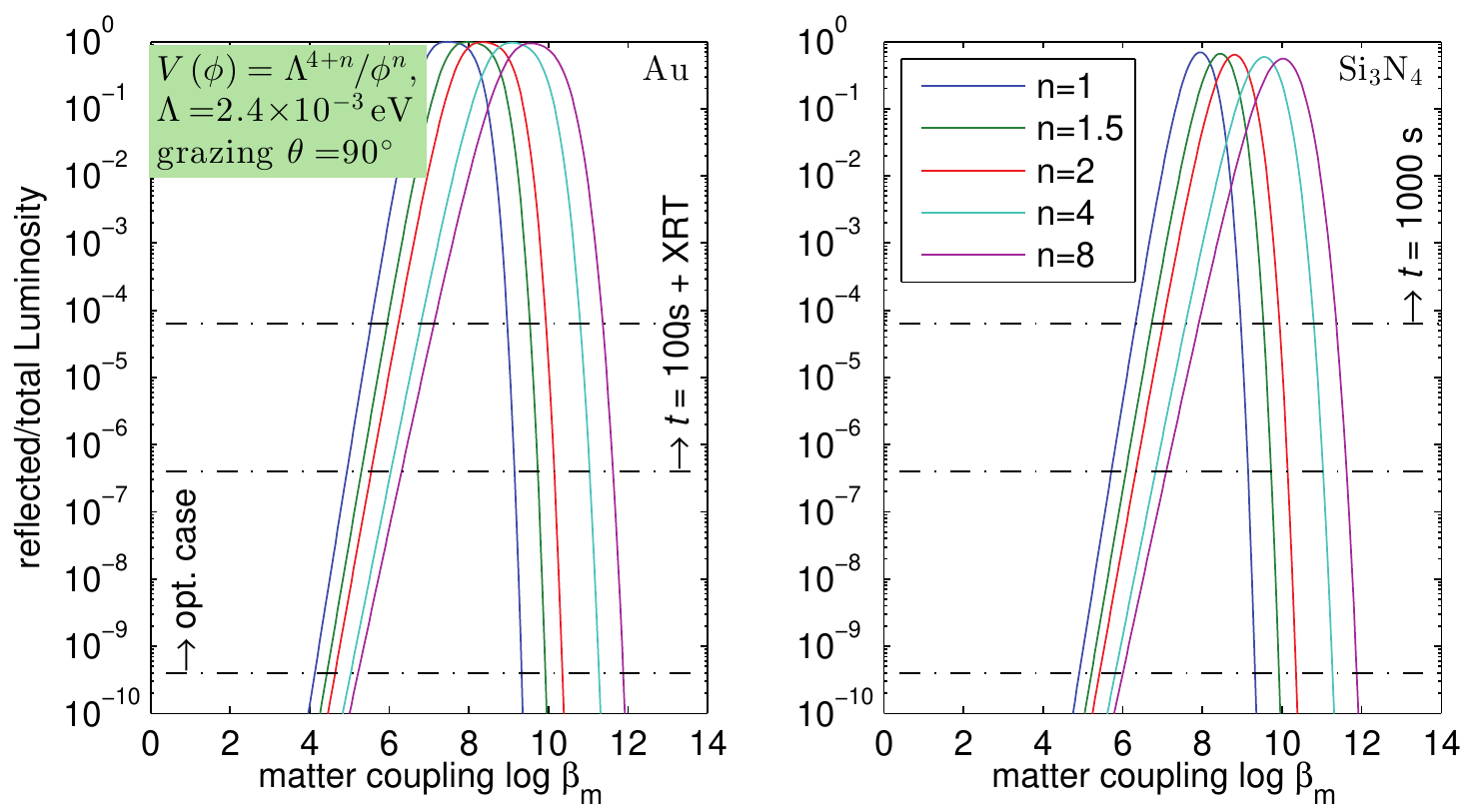}
	\caption{Fraction of the total chameleon flux reflected by the sensor membrane for a grazing angle of $5^{\circ}$ (left panels) and $90^{\circ}$ (right panels) with the chameleon potential mass scale fixed to the dark energy scale $\Lambda = 2.4\times10^{-3}\,$eV and assuming chameleon production only inside the tachocline. The 1st and 3rd panel from left show the fractions for the membrane coated with gold, and the 2nd and 4th panel for the bare membrane, respectively. The horizontal black dash-dotted lines show the minimum fraction needed to detect chameleons, assuming $L_{\text{cham}}/L_{\text{sol}} = 0.1$ and a measurement duration of 1000\,s without CAST's XRT, 100\,s with XRT. An optimal case with the membrane cooled to 0.3\,K, $t_{\text{meas}} = 100\,$s, a chopper system and using CAST's XRT is also shown. The corresponding value for $\beta_{\gamma}$ can be read off Figure \ref{fig:10percentandlumi}. $\beta_{\gamma}$ one order of magnitude smaller then corresponds to shifting the horizontal dash-dotted lines upwards by two order of magnitude since $\Phi_{\text{chameleon}}\propto \beta_{\gamma}^{2}$.}
	\label{fig:different_angles}  

	\includegraphics[width=.33\linewidth, trim=0cm 0cm 0cm 0cm, clip]{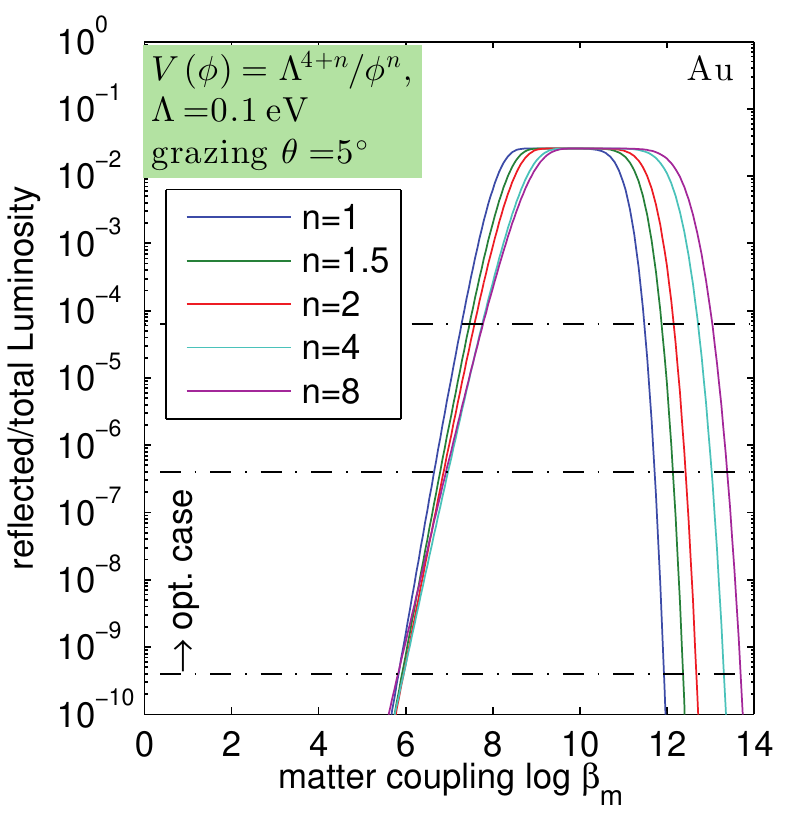}
	\includegraphics[width=.33\linewidth, trim=0cm 0cm 0cm 0cm, clip]{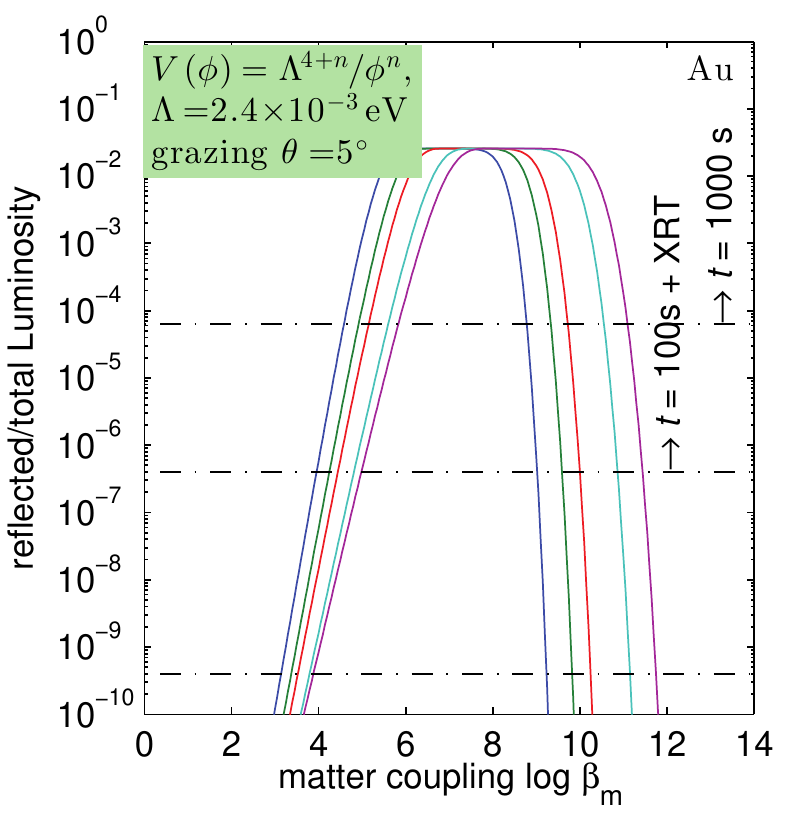}
	\includegraphics[width=.33\linewidth, trim=0cm 0cm 0cm 0cm, clip]{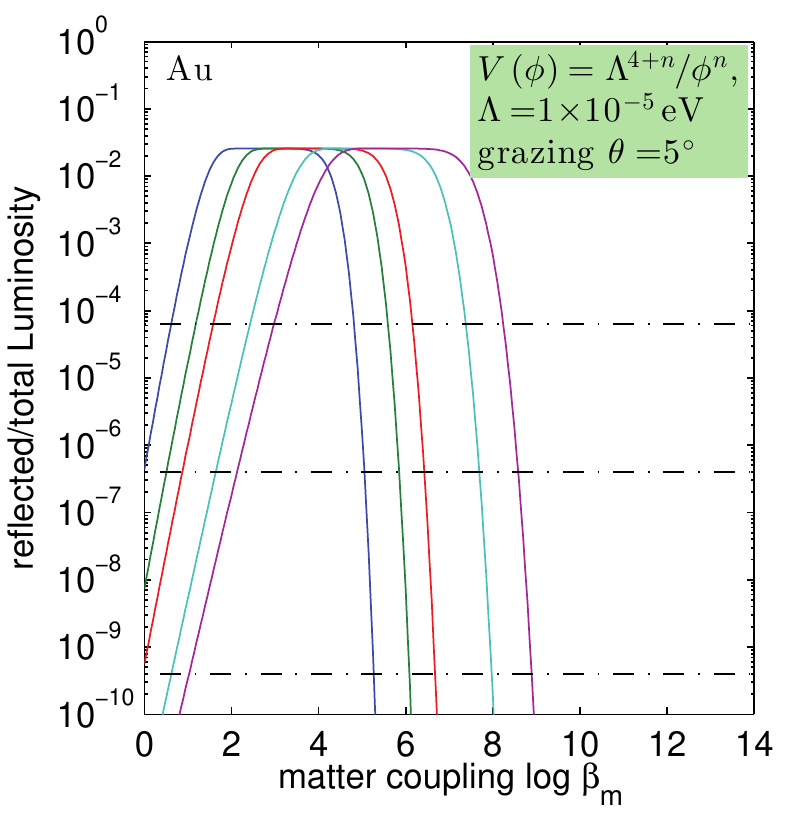}
	\caption{Fraction of the total chameleon flux reflected by the sensor membrane for a grazing angle of~$5^{\circ}$. The chameleon potential mass scale is fixed to 0.1\,eV (left), $2.4 \times 10^{-3}\,$eV, the dark energy scale, (middle) and $10^{-5}\,$eV (right) respectively. For further explanations, see the description of Fig. \ref{fig:different_angles}.}
	\label{fig:different_Massscales}

	\includegraphics[width=.33\linewidth]{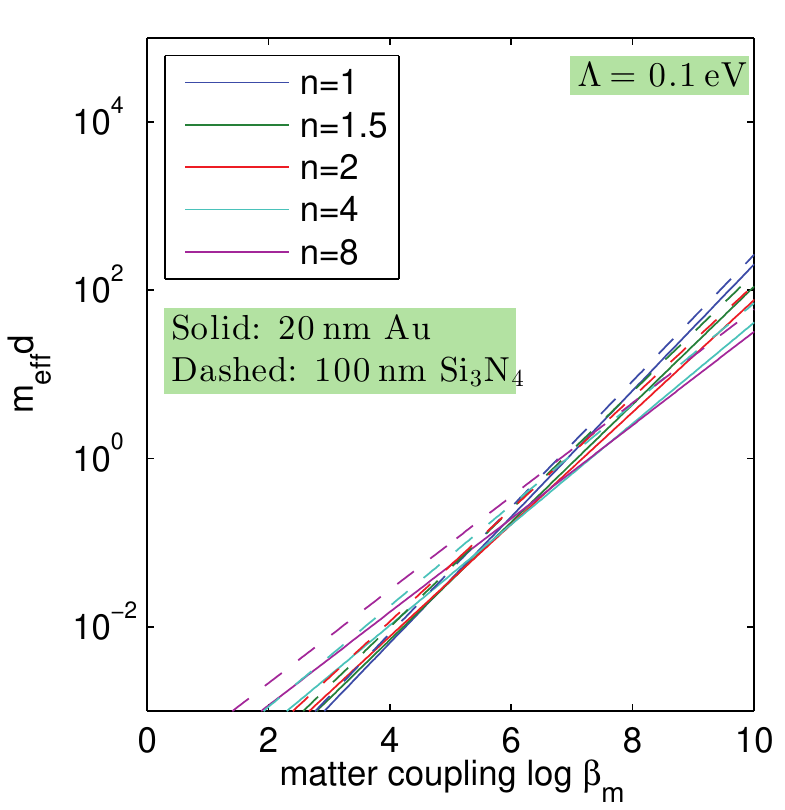}
	\includegraphics[width=.33\linewidth]{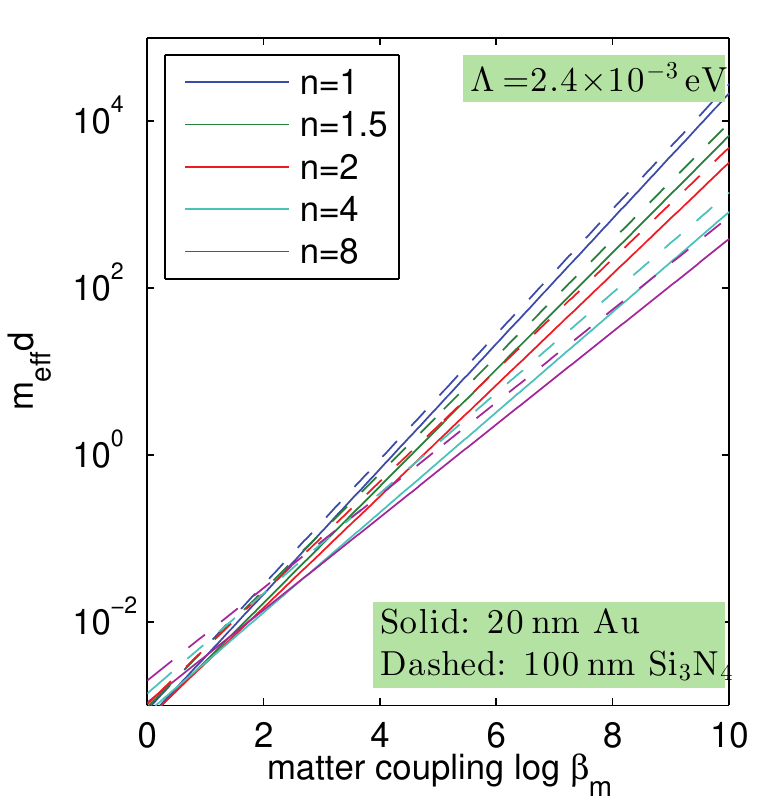}
	\includegraphics[width=.33\linewidth]{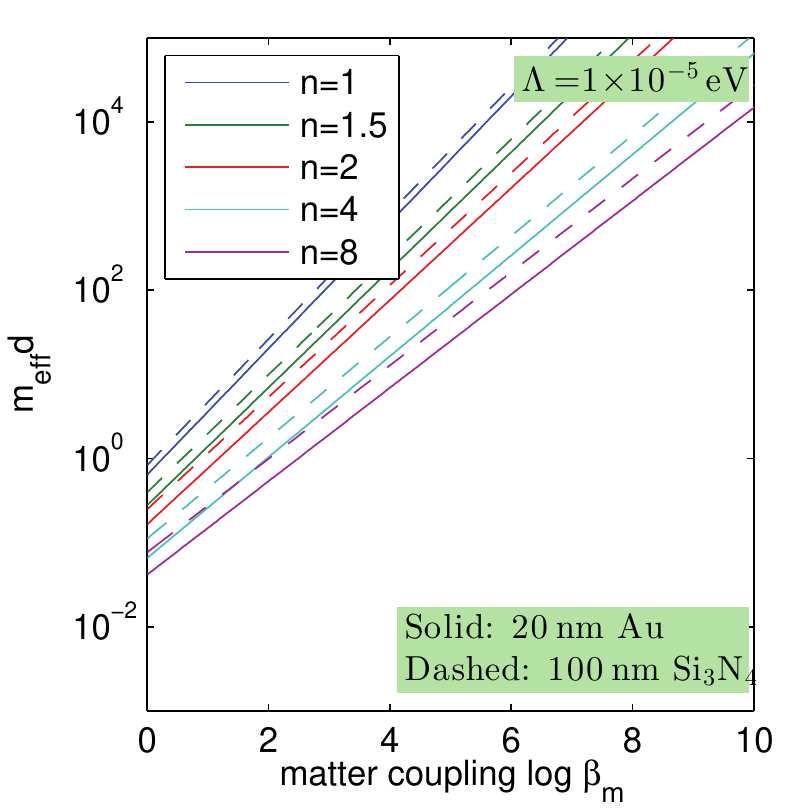}
	\caption{Effective mass in gold (membrane) $\times$ thickness of gold coating (membrane) vs. matter coupling for a chameleon model mass scale of 0.1\,eV (left), $2.4\times10^{-3}\,$eV, the dark energy scale, (middle) and $10^{-5}\,$eV (right). Only chameleons with $m_{\text{eff}} d \gtrsim 1$ will be reflected.}
	\label{fig:m_effd}
\end{figure}

\paragraph{Projected sensitivity}
 The sensitivity of KWISP to solar chameleons depends strongly on the chameleon model. The parameter space accessible by the proposed experiment is shown in Figures \ref{fig:different_angles},\ref{fig:different_Massscales}. Models where the reflected luminosity lies above the respective horizontal lines can be constrained. Since the detection principle relies on rapid scaling of the effective mass with the local matter density, the experiment will be most sensitive to strongly coupled chameleons, i.e. large matter couplings $\beta_m$ and small mass scales $\Lambda$. If the coupling becomes too strong, chameleons will not be able to reach the detector. Due to the shape of the solar chameleon spectrum, one can extend the sensitivity to smaller matter couplings $\beta_{m}$ by mounting the membrane at a small grazing angle with respect to the incoming chameleon flux (cf. Figure \ref{fig:different_angles}). Operating the sensor at a grazing angle of $5^{\circ}$ extends the sensitivity about one order of magnitude downwards in $\beta_m$ with respect to operating at $90^{\circ}$. The optimal angle depends on the parameters $n, \beta_m$, and $\Lambda$. Experimental feasibility needs to be taken into account. An angle of $\approx 5^{\circ}$ constitutes a good compromise between sensitivity to a broad range of chameleon models and experimental feasibility. 
 \par{} There are two potential quantum corrections which could affect detection: fragmentation and loop corrections to the potential.  Ref. \cite{Brax14} showed that proximity to a dense object suppresses fragmentation, since an increase in the effective mass raises the energy cost of producing a new chameleon particle.  Thus if fragmentation can be neglected in the chameleon population within the Sun, then further fragmentation due to scattering from the detector should be negligible. Meanwhile, quantum corrections modify the shape of the effective potential, making our tree-level mass calculations unreliable, when $m_\text{eff} > 0.007\,\text{eV}\left(\beta_m \times \rho/10\,\text{g.cm}^{-3}\right)^{1/3} \sim 1\,$eV in our range of interest. In other words, we do not know $m_\text{eff}$ and $V_\text{eff}$ very close to dense matter. However, our experiment is not a precision probe of the chameleon potential (as is, for example, a short-range fifth force experiment); it can only distinguish between particles which do and do not reflect. Furthermore, any residual effect of loop corrections to the potential can be mitigated by conducting the experiment at grazing incidence, where reflecting particles probe regions with lower $m_\text{eff}$ hence better-controlled quantum effects.
 \par{} The generation of large effective mass of the chameleons is a dynamical effect, hence, the membrane/gold coating has to be sufficiently thick for the chameleons to acquire large effective masses within. A complete treatment of this can be found in \cite{Mota07}. For our purposes a simplified consideration preserving the essential result suffices: we demand the chameleons' Compton wavelength $m_{\text{eff}}^{-1}$ in the membrane to be of the order of the thickness of the membrane/cold coating, respectively, or smaller (cf. Figure \ref{fig:m_effd}).
\par{} If we assume the chameleon mass scale to be the dark energy scale, $\Lambda = 2.4\times10^{-3}\,$eV, $L_{\text{cham}}/L_{\text{sol}} = 0.1$, and a first set-up using CAST's XRT and measuring for 100\,s, which is the time the XRT can focus solar chameleons without being moved, the experiment would already explore an uncharted area of parameter space of $\beta_m \sim 10^{4} \ldots 10^{12}$, depending on the power $n$ in the potential. The corresponding values of $\beta_{\gamma}$ can be read from Figure $\ref{fig:10percentandlumi}$. In an optimal case, one would cool the membrane below 0.3\,K and employ a chopper to modulate the chameleon flux. This extends the sensitivity to smaller matter couplings $\beta_m \gtrsim 10^{3}$. 
\par{}If we instead assume a much bigger mass scale $\Lambda = 0.1\,$eV, the experiment would only be sensitive to very strongly coupled chameleons, whilst in the case of a much smaller mass scale $\Lambda = 10^{-5}\,$eV, the experiment would be sensitive to weaker coupled chameleons, with the sensitivity to strongly coupled models diminished, since they could not be produced in the sun anymore.
\par{}If the magnetic field inside the sun is not confined to the tachocline but is linearily decreasing from the tachocline to the surface, the sensitivity will be extended to matter couplings $\beta_m$ one order of magnitude smaller, since production further outside enhances the chameleon flux at lower energies.
\par{}If we consider the case of materials denser than the vacuum window between the sun and the detector, e.g. lead shielding with $\rho_m = 11.3\,$g/cm$^{3}$, the sensitivity of the experiment remains unchanged on the low end of the $\beta_{m}$ space, whilst the sensitivity at the high $\beta_m$ side is decreased by roughly two orders of magnitude, given that the membrane is coated with gold or the bare membrane is mounted at a grazing angle smaller 20$^{\circ}$. We furthermore showed by comparing the chameleon's Compton wavelength with the thickness of the gold coating or membrane, respectively, that for chameleon-models within the sensitivity a 20\,nm thick gold coating or 100\,nm thick uncoated membrane suffices.

\paragraph{Conclusions}
We calculated the solar chameleon spectrum for a range of model parameters. We also showed that detecting the pressure caused by the reflection of solar chameleons from a micromembrane gives access to a large portion of the parameter space and that the sensitivity of the KWISP force sensor currently under development at INFN Trieste is sufficient to explore chameleon models with matter coupling $\beta_m \sim 10^{3} \ldots 10^{12}$ and photon coupling down to $\beta_{\gamma} \gtrsim 10^{7}$. Uncertainties due to quantum corrections to the chameleon have been discussed, but the literature is as yet inconclusive about the impact this might have on our sensitivity. However, such effects can be partially mitigated by mounting the sensor membrane at a small grazing angle with respect to the solar chameleon flux.
\par{}This apparatus is a unique pioneering effort in the field of experimental searches for dark energy. It will complement and complete different chameleon searches. A full comparison of the constraints from the different experiments is complicated by the wealth of chameleon models and results being often published only for a small subset of the parameterspace discussed in this work. Afterglow experiments \cite{GammeV-CHASE12} and previous searches at CAST looking for X-Ray photons from chameleon-photon conversion inside the CAST magnet are sensitive to the photon coupling and constraint the parameterspace to $\beta_{\gamma} \lesssim 10^{11}$ for a broad range of $n$ and $\beta_{m}$ \cite{GammeV-CHASE12}. Searches for the violation of Newtonian gravity \cite{EotWash07} are sensitive to weakly coupled chameleons. Limits have this far only been published for quartic models $n=-4$ and exclude $10^{-2} < \beta_m < 1$ for $10^{-7}\,\text{eV} < \Lambda < 2.4\times10^{-3}\,$eV \cite{EotWashRes}. Casimir force experiments are sensitive to strongly coupled models. As of yet, for $n>0$ only models with $\Lambda \gg 2.4\times10^{-3}\,$eV can be ruled out \cite{Casimir10}. Experiments with gravitationally trapped ultracold neutrons exclude $\beta_m > 10^{-9}$ for $\Lambda = 2.4\times10^{-3}\,$eV and a broad range of $n$, $\beta_{\gamma}$ \cite{qBounce14}. We showed, that the experiment proposed in this work will be sensitive to $10^{3} \lesssim \beta_m \lesssim 10^{10}$ and $\beta_{\gamma}$ as small as $10^{6}$ for $n>0$, $\Lambda = 2.4\times10^{-3}\,$eV. Even at its present sensitivity it will probe uncharted regions in the chameleon parameterspace.

\paragraph{Acknowledgements} This work has been partially supported also by grant 13.12.2.2.09 of the University of Rijeka.


\begin{thebibliography}{99}
\bibitem{Kho04-fields} J. Khoury and A. Weltman, PRL 93, 171104 (2004)
\bibitem{Kho04-cosmo} J. Khoury and A. Weltman, Phys. Rev. D 69, 044026 (2004)
\bibitem{Khoury13} J. Khoury, Class. Quantum Grav. 30 (2013), 214004
\bibitem{Brax11} P. Brax \emph{et al.}, Phys. Lett. B 699 (2011) 5
\bibitem{Hui10} L. Hui and A. Nicolis, PRL 105, 231101 (2010)
\bibitem{EotWashRes} A. Upadhye, Phys. Rev. D86, 102003 (2012)
\bibitem{Baker12} K. Baker \emph{et al.}, arXiv:1201.6508v1 (2012)
\bibitem{Brax10} P. Brax and K. Zioutas, Phys. Rev. D 82, 043007 (2010)
\bibitem{Brax12} P. Brax, A. Lindner, and K. Zioutas, Phys. Rev. D 85, 043014 (2012)
\bibitem{Cou03} S. Couvidat \emph{et al.}, ApJ 599, 1434 (2003)
\bibitem{Anita03} H.M. Anita, S.M. Chitre, and M.J. Thomson, A\&A 399, 329-336 (2003)
\bibitem{Burr09} C. Burrage, A. Davis, and D. Shaw, PRL 102, 201101 (2009)
\bibitem{Mit92} R. Mitalas and K. R. Sills, ApJ 401, 759 (1992)
\bibitem{Brax14} P. Brax and A. Upadhye, JCAP 1402 (2014) 018
\bibitem{Erickcek14} A. L. Erickcek \emph{et al.}, PRL 110, 171101 (2013) and Phys. Rev. D 89, 084074 (2014)
\bibitem{Cantatore1995} G. Cantatore \emph{et al.}, Rev. Sci. Instrum. 66, 2785 (1995)
\bibitem{Karuza12} M. Karuza \emph{et al.}, New J. of Phys 14 095015 (2012)
\bibitem{Yu12} P.-L. Yu, T. P. Purdy, and C. A. Regal, PRL 108, 083603 (2012)
\bibitem{Gardikiotis} A. Gardikiotis, private communication.
\bibitem{Baker12-2} K. Baker \emph{et al.}, arXiv:1201.0079v1 (2012) 
\bibitem{Xu13} X. Xu, J.J. Taylor, arXiv:1303.7469v1
\bibitem{Mota07} D.F. Mota and D.J. Shaw, Phys. Rev. D 75, 063501 (2007)
\bibitem{GammeV-CHASE12} A. Upadhye, J. H. Steffen, and A. S. Chou, Phys. Rev. D 86, 035006 (2012)
\bibitem{EotWash07} E. G. Adelberger \emph{et al.}, PRL 98, 131104 (2007)
\bibitem{Casimir10} P. Brax \emph{et al}., Phys. Rev. D 76, 124034 (2007) and PRL 104, 241101 (2010) 
\bibitem{qBounce14} T. Jenke \emph{et al.}, PRL 112, 151105 (2014)
\end{thebibliography}
\end{document}